**Selective Chemical Modification of Graphene surfaces: Distinction between Single and Bilayer Graphene \*\***

*Fabian M. Koehler, Arnhild Jacobsen, Klaus Ensslin, Christoph Stampfer and Wendelin J. Stark* [*]

[*]     Prof. Wendelin J. Stark, Fabian M. Koehler
Institute for Chemical and Bioengineering
ETH Zurich, CH-8093 Zurich (Switzerland)
E-mail: wendelin.stark@chem.ethz.ch

        Prof. Klaus Ensslin, Prof. Christoph Stampfer, Arnhild Jacobsen
Solid State Physics Laboratory
ETH Zurich, CH-8093 Zurich (Switzerland)

        Prof. Christoph Stampfer
JARA-FIT and II. Institute of Physics
RWTH Aachen, 52074 Aachen (Germany)

Keywords: Covalent functionalization, Molecular electronics, Raman spectroscopy, Reaction mechanisms, Electronic structure

**Abstract**

Graphene modifications with oxygen or hydrogen are well known in contrast to carbon attachment to the graphene lattice. The chemical modification of graphene sheets with aromatic diazonium ions (carbon attachment) is analyzed by confocal Raman spectroscopy. The temporal and spatial evolution of surface adsorbed species allowed accurate tracking of the chemical reaction and identification of intermediates. The controlled transformation of $sp^2$ to $sp^3$ carbon proceeds in two separate steps. The presented derivatization is selective for single layer graphene and allows controlled transformation of adsorbed diazonium reagents into covalently bound surface derivatives with enhanced reactivity at the edge of single layer graphene. On bi-layer graphene the derivatization proceeds to an adsorbed intermediate without further reaction to form a covalent attachment on the carbon surface.

## 1. Introduction

Recently it has become possible to fabricate graphene on large areas.[1,2] In order to control the electronic properties of this fascinating material it is desirable to chemically modify graphene surfaces. One example is the oxidation to graphene oxide[3,4], i.e. randomly attaching oxygen species to the graphene lattice by breaking the C-C π- and σ-bonds which results in a breakdown of conductivity.[5] A more sensitive method has been described by Elias et. al.[6] and proceeds by hydrogenation of the $sp^2$ carbon atoms at least on one graphene side by breaking the C-C π bonds. This allows to preserve the crystalline order of the carbon lattice, but leads to re-hybridization of the carbon atoms from a $sp^2$ to a distorted $sp^3$ state, which induces the formation of a band gap.[6] The hydrogenation is reversible through heating and proceeds by de-hydrogenation of the graphane to graphene. As an extension of the hydrogenation concept (C + H = CH), the covalent attachment of small organic molecules [7,8] now enables access to a library of functional groups on graphene surfaces. Such chemical functionalization may either be used to change the electronic properties or the functional groups providing an anchor to sophisticated organic chemistry.[9] This could allow the use of graphene as a target sensitive sensor or a basis to couple graphene to molecular electronics.

Recently we have used a wet-chemical method [7,10] to demonstrate the above concepts of chemically modifying graphene by covalently attaching organic moieties in a pattern-controlled way (lithography) onto a graphene model surface (highly oriented pyrolytic graphite (HOPG)).[11] More specifically, attachment of p-substituted benzene rings perpendicular to the graphene plain allowed to influence directly the surface potential by correctly choosing the benzene substituents. Electron donating groups (e.g. methoxy: $CH_3O$-) push electrons into the benzene substituent and hence locally into the graphene sheet. By analogy, electron withdrawing groups (e.g. nitro: $NO_2$-) can be used to locally remove electrons (3D silicon analogy: n- or p-doping). These subtle changes in the electron density affected the surface potential by about 100 mV. Similar chemistry has recently also been applied to few layer epitaxial graphene.[12,13] A theoretical investigation of the spatial distribution of the attached benzene rings indicates their potential crystalline order on the graphene surface. Using a temperature dependent transport experiment Bekyarova et al. confirmed the different behavior of functionalized versus native graphene, as a result of the induced gap.[12,13] In contrast, Farmer et al.[14] showed that the above diazonium chemistry resulted only in p-doping on single layer graphene and attributed this limitation to a non-covalent charge transfer complex formed on the graphene surface.

In this work, we apply functionalization chemistry to single layer and bi-layer graphene sheets and confirm the chemical introduction of molecules by atomic force microscopy (AFM) and scanning confocal Raman spectroscopy which is the spectroscopic method of choice for rapidly characterizing graphene, both for the number of graphene layers (width of 2D line)[15,16], graphene flake quality (D line) and the degree of doping (i.e. shift of Fermi energy, position of G line and 2D line)[16,17]. We thereby demonstrate the controlled introduction of defects through chemical modification. In addition, confocal Raman spectroscopy allows to distinguish between $sp^3$ and $sp^2$ bonds of hybridized carbon. [18]

## 2. Results and Discussion

There are several chemical reaction pathways how the above reagent can react with the carbon surface. In parallel, we must consider the possibility that the reagent or an intermediate not only takes part in the chemical reaction, but also just adsorbs on the surface itself. **Figure 1** schematically shows the two interactions (chemical reactions or physisorption) on single and bi-layer graphene. The 4-nitrobenzene-diazonium-tetrafluoroborate (NBD, **1**) can decay to a reactive intermediate (chemical reaction), or decompose to nitrobenzene (**6**). Both nitrobenzene or the starting compound may adsorb on graphene.[19] To differentiate between these two mechanisms, nitrobenzene was adsorbed on the graphene flakes in a separate control experiment and analyzed with the same procedures as for the chemically functionalized flakes.

**Figure 2** shows Raman spectra between adsorbed (bottom pattern; graphene before and after physisorption of nitrobenzene) and chemically modified graphene (top pattern, graphene before and after exposure to the reagent (Fig. 1, NBD, **1**)). While nitrobenzene adsorption did not provoke any obvious changes in the Raman spectra, it induced a right shift of the G line (1586 to 1589 $cm^{-1}$). This can be attributed to a doping effect, in agreement with the adsorption of larger aromatic molecules as reported by Donget al.[20] In line with the fact that an adsorbed molecule does not change the underlying crystalline structure of graphene, no signal has been observed in the range of the D-line (characteristic for defects).

Comparing adsorption (Fig. 2, bottom) with functionalization (Fig. 2, top) reveals clear differences in the spectra. After 10 minutes exposure to the diazonium reagent (**1**) the D-(1340 $cm^{-1}$) peak strongly increased. The signal between the D and G line (1370 to 1540 $cm^{-1}$) is also slightly elevated, which has been interpreted as a sign of $sp^3$ carbon formation[18] as observed in thin diamond (bulk diamond: pure $sp^3$ carbon) films.[21] The re-hybridization of the carbon atoms ($sp^2$ to $sp^3$) in graphene has been shown for the chemically more simple graphane (C + H = CH) by a similar Raman study-based argument.[6] In addition, Bekyarova et al.[12] have investigated this transition using diazonium chemistry functionalized graphene by X-ray photoelectron spectroscopy (XPS) and analyzing the $C_{2S}$ band.

The distinction between single and bi-layer graphene made in figure 1 is illustrated in the following experiments. **Figure 3** shows a representative flake, where the Raman maps of the FWHM of the 2D line (Fig. 3a) and the G line intensity (Fig. 3d) illustrate the position of single layer (1L), bi-layer graphene (2L) and few-layer graphene. In addition AFM topography images show the flake before and after the chemical treatment. The chemically induced change resulted in an increased thickness determined from the height profiles of the single layer graphene flake from $2.0 \pm 0.4$ nm to $4.5 \pm 0.5$ nm. The temporal evolution of the chemically induced changes of this flake can be followed in the time sequence of the two dimensional Raman maps (**figure 4**). The D line intensity explicitly shows that the diazonium ion first produced defects predominantly in single layer graphene (more clearly shown in **figure 5**), with pronounced intensity at the edges. The evolution of defects clearly differentiates between single and double layer areas.

The Raman spectra in figure 5 highlight the absence of the D-line (i.e. defects) in functionalized bi-layer graphene. The D-line grows for single layer graphene in time, and it is even stronger at the edge of single layer graphene. The D/G intensity ratio is more than 50% higher at the edge than at the bulk of a single layer, indicating a higher degree of defect formation. The adjacent peaks at 1400 and 1440 cm$^{-1}$ can be attributed to adsorbed diazonium ion molecules.[22] They occur after about 20 minutes on single layer graphene (bi-layer: 40 minutes) and are thought to build a charge transfer complex with the graphene surface.[14] As the diazonium peaks occur dominantly after 20 minutes on single layer graphene, it seems that a different chemical reaction happened on the surface. This result can best be interpreted, when assuming that in the beginning only defects are produced (molecules attached to the surface) on the single layer. This could mean that species (**3**) is formed on the surface at positions, where it is energetically favorable to rearrange the carbon lattice to form a local $sp^3$ carbon geometry. After all these sites have reacted, (**2**) seems to stay adsorbed on the carbon lattice as the formation of sp3 carbon atoms has a higher activation energy and hence the diazonium peaks are visible. The same holds true for bi-layer graphene. However, the covalent attachment (**5**) does not occur and only the adsorbed charge transfer complex (**4**) can be seen in the Raman spectra. The diazonium ion peaks are of weak intensity at the edge of the graphene layer and progressively disappear on the single layer flake as it becomes more functionalized (i.e. covalent functionalization). This process is best visible in the maps of the D peak (figure 4) where the edge appears in bright yellow (very high intensity) and, with time, this region grows into the middle of the single layer graphene. After 80 minutes, the single layer area has a high D line intensity with similar intensity and all diazonium peaks have vanished. The increased background signal between D and G line (1370 to 1540 $cm^{-1}$) is another indicator for $sp^3$ carbon formation. Again this feature is more pronounced at the edge of the single layer rather than on the bulk single layer. The bi-layer has only a slightly increased background signal. In addition the formation of a structural change can be seen in the growth of the D' peak (1620 $cm^{-1}$), which is present on the single layer edge and gives rise to the shoulder of the G line in the spectra of single layer graphene.

These observations stay in full agreement with the chemical reaction pathway given in figure 1. The reagent first has to adsorb and then to react, as supported by the Raman band observed here. Both single layer regions at the bottom and at the right of the flake have a similar rate of edge growth. The constrained, small single layer band (middle, Fig. 3a) which lies between the two double layer regions behaved differently. Similar to the double layer area, it developed a slowly growing edge.

**Figure 6** shows selected data points extracted from the Raman spectra (D/G intensity ratio, G line position and D line intensity). During chemical functionalization, the D/G intensity ratio of the single layer and its edge are significantly higher and increase faster than for double layer graphene. The rate constants shown in figure 6a provide a quantitative measure for the different reactivities of the diazonium species on the edge of single layer, as well as bulk of single layer and bi-layer graphene. The relatively small rate constant for bi-layer graphene reveals that here the reaction does not proceed to the covalently bound species (**5**). This observation confirms the different ability to react (rearrangement of the carbon lattice; accommodate strain through the altered geometry at the reaction sites) preferably at the edge (easy to detach or shift the edge region) versus single layer areas.

A second, clear difference between single- and bi-layer in the functionalization experiments is the shift of the G-line to higher energies followed by a slow decrease to lower energies due to a change of charge carrier concentrations in graphene (doping effect, figure 5b).[23,24] The distinction between p- or n-doping can hardly be made from Raman data only. Our previous Kelvin force microscopy (KFM) study on chemical functionalization of graphene model surfaces using a series of substituted benzene moieties (different reagents, Figure 1, exchange of $NO_2$ by COOH, $SO_3H$ or $OCH_3$) suggests p-doping since nitrobenzene is a strong electron withdrawing substituent as extensively shown in the so called Hammett correlations in organic chemistry [25,26]. The change of the Raman shift in single layer graphene of about 4 cm$^{-1}$ corresponds to a change in the charge carriers of about n ≈ – 5 x 10$^{12}$ cm$^{-2}$, whereas for bi-layer graphene the doping is not significant.

Combining both Raman results shows that the aryl diazonium ion preferably binds to single layer graphene and only weakly (i.e. later) to bi-layer areas. Hence, one can chemically differentiate between these two systems. In addition, the defect peak is never visible on bi-layer, confirming that no defect formation takes place in the bi-layer area. The non-covalent diazonium charge transfer complex (**4,** 1400 and 1440 cm$^{-1}$)[22] remains adsorbed without decaying nor forming a covalent carbon-carbon bond (**5**) to bi-layer graphene. This different reactivity of single and bi-layer graphene is in line with observations on the hydrogenation of graphene to graphane. There, single side hydrogenation is assumed and surface ripples of single layer graphene[27] are assumed to reduce the activation energy for the formation of sp3 carbon atoms on a surface[6]. Bi-layer graphene is more crystalline[6,28] (less ripples) and the flexibility to accommodate a local sp$^3$ geometry at the surface is reduced due to the underlying carbon lattice thus offering an explanation for the different reactivity. The reduced activity of the small single layer graphene strip (Figure 3(a) and 4, rigidly embedded between two bi-layer areas) again confirms the role of graphene sheet flexibility. Moreover, it couples the chemical reactivity to a macroscopic geometric effect. In addition the edges are assumed to have a random structure of sp$^2$ and sp$^3$ carbon atoms, which enhance the decay of the adsorbed diazonium ion to the covalently bound reagent (3), as the energy for distorting the graphene and building a $sp^3$ carbon atom is smaller than on single layer graphene itself.

After functionalization a strong D line occurs on the single layer graphene areas indicating a partial change of the electronic properties of the carbon structure and also the introduction of defects to the surface. The initial presence of the diazonium ion peaks are an indication of the non covalent attachment of the reagent to the $sp^2$ carbon atoms, which would result only in a doping effect shown by Farmer et al.[14] and not in the opening of a gap due to $sp^3$ hybridization. Following the two step mechanism suggested above (Figure 1), the diazonium ion progressively vanishes and is replaced with covalently bound species. From topography images obvious defects were not observable, however they showed an increased height after functionalization, which is a result of the attached nitrobenzene groups standing perpendicularly to the graphene surface.[11] This observation and the possibility to distinguish between adsorbed and chemically reacted species are in agreement with the parallel, subsequently observed formation of defects. For longer exposure, the defect concentration (D-line) on single layer graphene increases with time until saturation occurs within about 1 hour. Chemical investigations have shown that further, more prolonged exposure to the diazonium reagent results in an oligo- or polymerization- type reaction with currently unknown structure.[11]

## 3. Conclusions

The presented method allows the controlled introduction of defects selectively into single layer graphene. Differentiation between single and bi-layer graphene may assist device fabrication processes where high quality layers of a specific type are required. The possibility to combine the above chemical derivatisation with classical lithography makes such methods amenable to conventional semiconductor manufacturing processing. In contrast to H-atom attachment for graphane, derivatisation provides a chemically reactive and versatile set of anchors to connect graphene to molecular electronics.

## 4. Experimental Section

Single layer graphene was prepared by scotch tape method of natural graphite flakes on Si/SiO$_2$ chips after Novoselov et al.[29] and identified by optical microscopy and characterized by Raman and AFM. The functionalization (attachment of perpendicularly oriented benzene rings, transition from sp$^2$ to sp$^3$ hybridization; planar to tetrahedral carbon geometry) was carried out by immersing the chip into 25 ml of a 20 mmol/l solution, prepared by weighing 120 mg 4-nitrobenzene-diazonium-tetrafluoroborat (Aldrich, 98%) in 25 g water (Millipore, resistance: 18 MOhm cm). The functionalization was carried out at room temperature (298 K) and the duration was varied in a series of experiments detailed below. After functionalization, the chips were cleaned once in isopropanol (Fluka, 99.8 %, ACS, 1 min), twice in water (Millipore, 1 min), again in isopropanol (Fluka, 99.8 %, ACS, 1 min) and dried with nitrogen (Pan Gas, 5.0). After each functionalization step, the graphene flakes were thoroughly analyzed by Raman spectroscopy to monitor the chemically induced changes.

For the nitrobenzene adsorption a graphene flake on a Si/SiO$_2$ chip was immersed in 20 ml of nitrobenzene for 10 minutes in the dark. Afterwards the chip was cleaned in isopropanol (Fluka, 99.8 %, ACS, 1 min) and dried with nitrogen (Pan Gas, 5.0).

**Acknowledgements**

The authors would like to thank Prof. Christofer Hierold for access to the confocal Raman microscope and Dr. Konstantin Novoselov for helpful advices. Financial support by the ETH Zurich (TH 02-07-3) and the Swiss National Science Foundation (SNF200021-116123) are gratefully acknowledged.

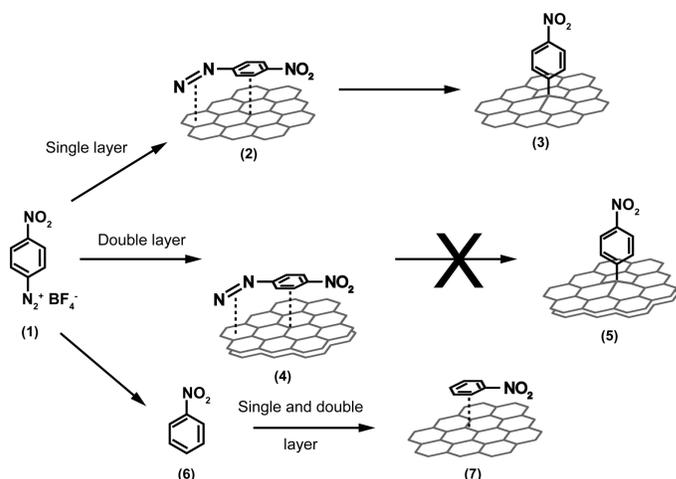

**Figure 1.** Schematic representation of the proposed reaction pathways on single and bi-layer graphene. The Diazonium reagent (1) presumably adsorbs on single-(2) or bi-layer graphene (4) and decomposes in a second step to form a covalently bound nitrobenzene moiety on the graphene surface (3). On bi-layer this decomposition of the diazonium group does not take place. (1) can additionally decay to nitrobenzene (6), which can only adsorb on the graphene surfaces, however no reaction with the graphene takes place.

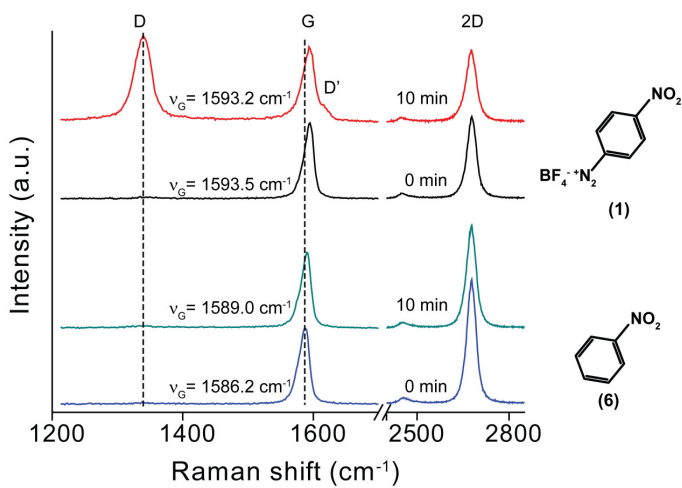

**Figure 2.** Distinguishing the two pathways (reaction and adsorption) by the same experiment with nitrobenzendiazonium tetrafluoroborat (top) and with nitrobenzene (bottom). The adsorption of nitrobenzene shows a shift of the G line after 10 minutes, corresponding to a doping effect, which also explains the smaller 2D line. With reagent (1) a strong D line and D' line is observable after 10 minutes, indicating the formation of defects in the graphene surface.

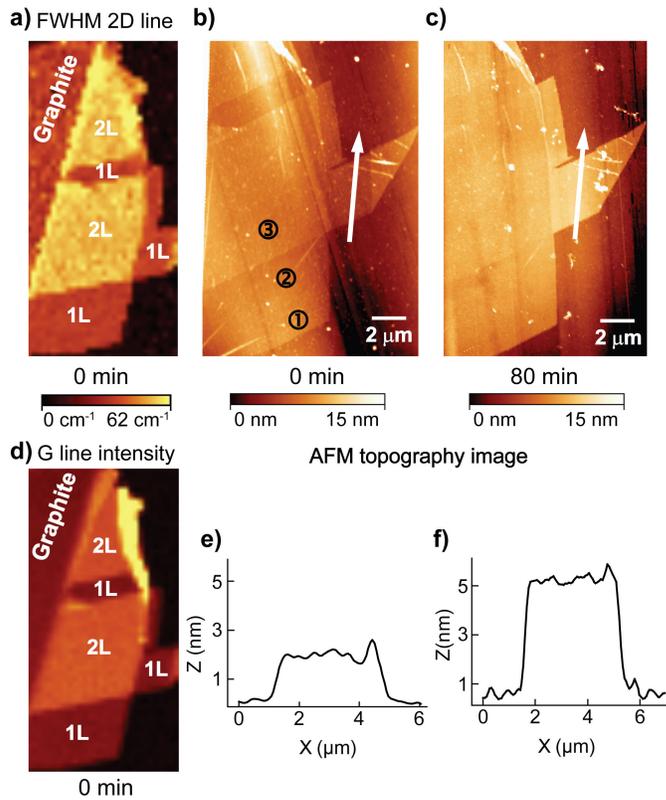

**Figure 3.** (a) Raman image of the full width at half maximum (FWHM) of the 2D line from a representative flake composed of single layer (1L), double layer (2L) and graphite. (b) and (c) AFM topography images before (0 min) and after (80 min) the functionalization experiment. The numbers in (b) indicate the position of the Raman spectra obtained for figure 4. 1) edge of single layer, 2) single layer and 3) bi.layer graphene. (d) Raman map of the G line intensity. (e) and (f) height profile along the white arrow in (b) respectively (c).

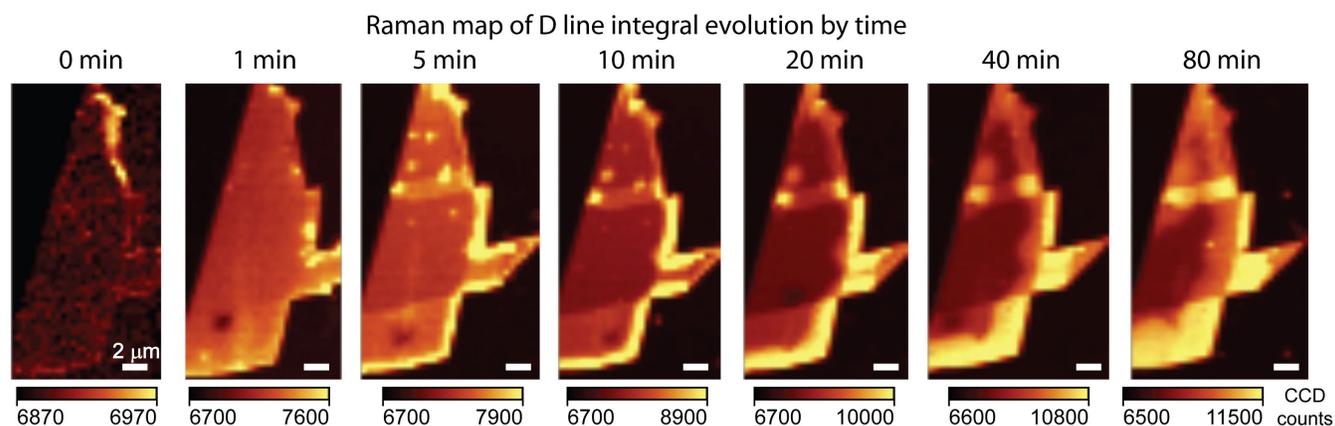

**Figure 4.** Raman images of the area under the D line of the graphene flake. After 5 minutes a clear differentiation of single (lighter, higher integral) and bi-layer graphene (darker, smaller integral) is possible. Additionally the edge of single layer graphene is now visible, highlighting the different rate constants of the reaction. After 40 minutes one can also see a difference in the reactivities of the small single layer embedded between the two bi-layer regions at the top of the flake compared to the single layer region at the bottom. This highlights that the reactions need a certain flexibility of the graphene flake to covalently bind a nitrobenzene moiety.

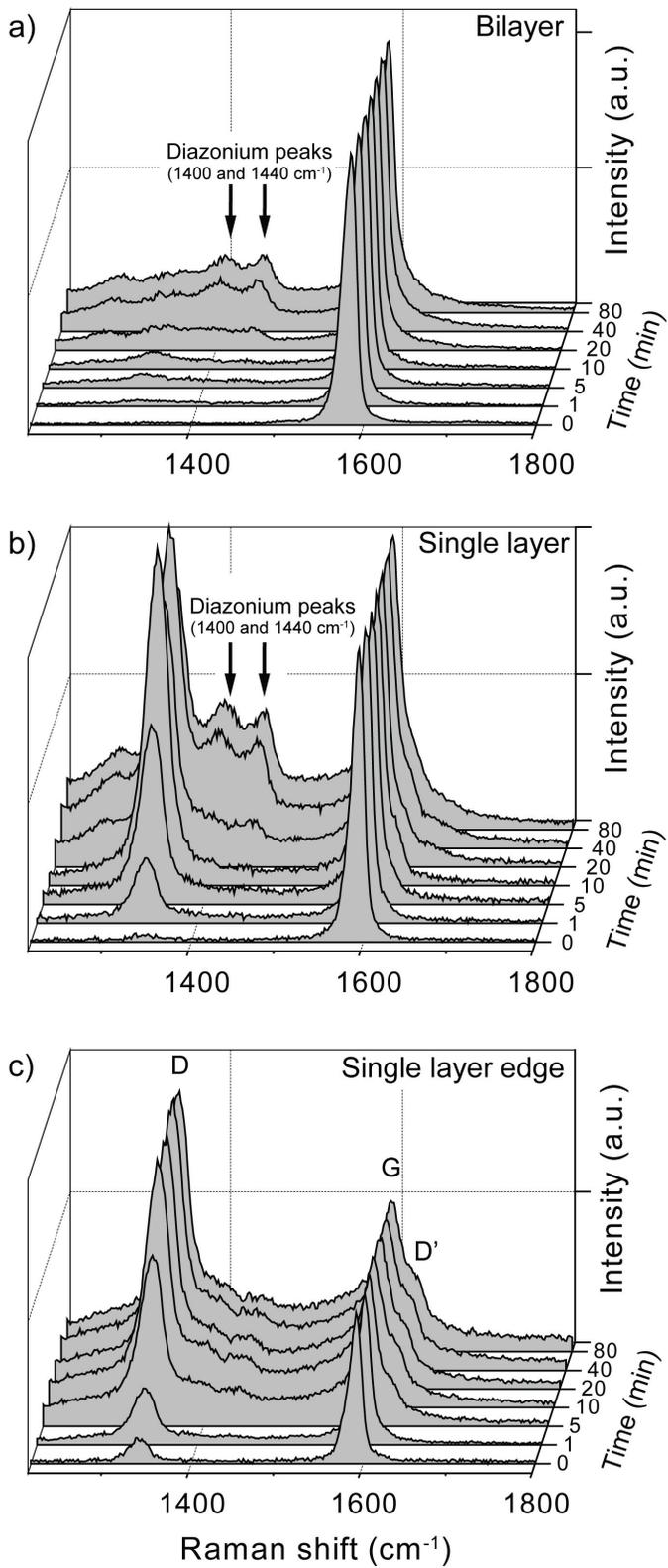

**Figure 5.** Raman spectra of the D and G line region during the experiments. (a) double layer graphene, (b) single layer graphene and (c) edge of single layer graphene. The absence of the D – line in (a) shows that no reaction took place, whereas the diazonium peaks indicate the presence of an adsorbed species on the surface. In the single layer (b) both species are present (adsorbed diazonium ion and covalently bound nitrobenzene) and on the edge only the covalently bound species is seen in the Raman spectra.

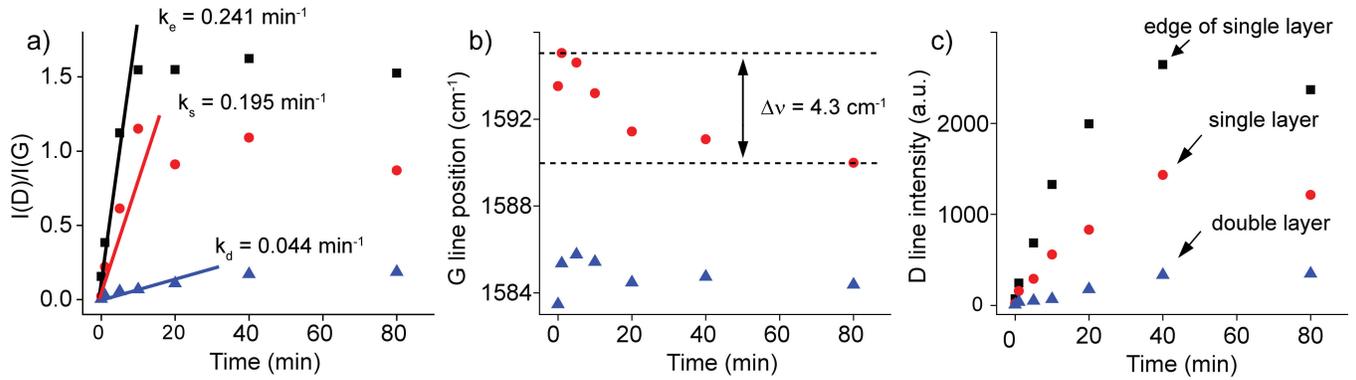

**Figure 6.** (a) Ratio of D line intensity over G line intensity for double layer (triangle), single layer (circles) and the edge of single layer (rectangles). From these data the rate constants were extrapolated, to quantify the different reactivities. (b) Position of G line is a measure of the doping level of graphene. Here the single layer experiences a change of charge carrier of about $n \approx -5 \times 10^{12}$ cm$^{-2}$ ($\Delta v = 4.3$ cm$^{-1}$) and (c) D line intensity at the specified times.